# Study of variations as desired-relative (δ), rather than absolute, differences: falsification of the purpose of achieving source-representative and closely comparable lab-results

*B. P. Datta* (email: bibek@vecc.gov.in)
Radiochemistry Laboratory, Variable Energy Cyclotron Centre, 1/AF Bidhan Nagar, Kolkata 700 064, India

**ABSTRACT**

Recently (arXiv:1101.0973), it has been pointed out by us that the possible variation in any source (***S***) specific elemental isotopic (viz. $^{2}$H/$^{1}$H) abundance ratio $^{S}\boldsymbol{R}$ can more accurately be assessed by its absolute estimate $^{S}\boldsymbol{r}$ (viz. as "**(**$^{S}\boldsymbol{r}$ – $^{D}\boldsymbol{R}$**)**", with ***D*** as a standard-source) than by either corresponding measured-relative ($^{S/W}\boldsymbol{\delta}$) estimate "**([**$^{S}\boldsymbol{r}$/$^{W}\boldsymbol{r}$**] – 1)**" or **δ**-scale-converted-relative ($^{S/D}\boldsymbol{\delta}$) estimate "**([**$^{S}\boldsymbol{r}$/$^{D}\boldsymbol{R}$**] – 1)**". Here, we present the fundamentals behind scale-conversion, thereby enabling to understand why at all "$^{S}\boldsymbol{r}$" should be the source- and/ or variation-characterizing key, i.e. why different lab-specific results should be more closely comparable as absolute estimates ($^{S}\boldsymbol{r}^{\mathrm{Lab1}}$, $^{S}\boldsymbol{r}^{\mathrm{Lab2}}$ **…**) than as desired-relative ($^{S/D}\boldsymbol{\delta}^{\mathrm{Lab1}}$, $^{S/D}\boldsymbol{\delta}^{\mathrm{Lab2}}$ **…**) estimates. Further, the study clarifies that**: (i)** the **δ**-scale-conversion "$^{S/W}\boldsymbol{\delta} \rightarrow {}^{S/D}\boldsymbol{\delta}$" (even with the aid of calibrated auxiliary-reference-standard(s) ***Ai(s)*** "$^{S/W}\boldsymbol{\delta} \xrightarrow{Ai(s)} {}^{S/D}\boldsymbol{\delta}$") cannot make the estimates (as $^{S/D}\boldsymbol{\delta}$, and thus $^{S}\boldsymbol{r}$) free of the measurement-reference ***W***; **(ii)** the employing of (increasing number of) ***Ai***-standards should cause the estimates to be rather (increasingly) inaccurate and, additionally, ***Ai(s)***-specific; and **(iii)** the $^{S/D}\boldsymbol{\delta}$-estimate may, specifically if ***S*** happens to be very close to ***D*** in isotopic composition (IC), even *misrepresent* ***S***; but the corresponding "$^{S}\boldsymbol{r}$" should be very *accurate*. However, for ***S*** and ***W*** to be increasingly closer in IC, the $^{S/D}\boldsymbol{\delta}$-estimate and also $^{S}\boldsymbol{r}$ are shown to be increasingly accurate, irrespective of whether the $^{S/W}\boldsymbol{\delta}$-measurement accuracy could thus be improved or not. Clearly, improvement in measurement-accuracy should ensure additional accuracy in results.

## 1. INTRODUCTION

Isotopic composition (IC) of any element, specifically lighter, is variation prone for geo, bio, environmental **…** changes. Thus, variations-in-ICs of stable lighter elements (as a function of sample-source/ time **…**) are themselves used as the tools for diagnosing the causes of IC-variations, origins-of-sample-sources, and so. The (variation of) IC in any sample ***S*** is usually measured as a "**δ**"-variable[1]**:** $^{S/W}\delta = [(^{S}R/^{W}R) - 1]$; with $^{S}R$ as any sample-isotopic [e.g. $^{2}H/^{1}H$] abundance ratio in question, and $^{W}R$ as the corresponding ratio in a working-lab-reference ***W***. The **δ**-measurement-technique is rather known as the isotope ratio mass spectrometry (IRMS); i.e. though the IRMS is supplemented from time to time in different ways[1-6], and even lately by the laser mass spectrometry[6-8]. However, lab specific choice of ***W*** should cause different lab-results to be incomparable. Thus, any (species-specific) result is evaluated with reference to a (corresponding) recommended[9,10] reference-standard (***D***). In other words, the *desired* variable is generally considered to be "$^{S/D}\delta$" [with**:** $^{S/D}\delta = ([^{S}R/^{D}R] - 1)$], but without examining whether the **δ**-scale-conversion**:** $^{S/W}\delta \rightarrow {}^{S/D}\delta$ can only remove the barrier for, or really ensure accurate, inter-comparison of lab-results ($^{S/D}\delta_{Lab1}$, $^{S/D}\delta_{Lab2}$ **…**). We are, therefore, concerned here.

The benefit of performing the measurement as "$^{S/W}\delta$" is that the effects of (uncorrectable) technical-biases should largely be eliminated. However, by an *estimate* (***r***) of any isotopic-abundance-ratio "***R***", it should mean**:** $R = (r + \Delta)$, or**:** $R = (r \pm u)$, with Δ as [true] measurement-error and ***u*** as uncertainty[11]. Therefore, the ratio "$^{S}r/^{W}r$" (of even ratios "$^{S}r \pm {}^{S}u$" and "$^{W}r \pm {}^{W}u$, which are acquired under *identical* possible experimental conditions) should[12], though sometimes [when their true-measurement-errors, $^{S}\Delta$ and $^{W}\Delta$, might equal to one another] represent the true-ratio "$^{S}R/^{W}R$", generally be more erroneous than either the estimate "$^{S}r$" or "${}^{W}\boldsymbol{r}$". That is[12]**:** (${}^{S}\boldsymbol{R}/{}^{W}\boldsymbol{R}$) = ([${}^{S}\boldsymbol{r}/{}^{W}\boldsymbol{r}$] ± [${}^{S}\boldsymbol{u}$ + ${}^{W}\boldsymbol{u}$]); and is because that ${}^{S}\boldsymbol{\Delta}$ and ${}^{W}\boldsymbol{\Delta}$ may, even in *direction*, differ from one another. Essentially, any measured-data as even ${}^{S/W}\boldsymbol{\delta}$ could be at variance (±${}^{S/W}\boldsymbol{u}$) over corresponding [unknown] true value. Thus, the conversion**:** (${}^{S/W}\boldsymbol{\delta}$ ± ${}^{S/W}\boldsymbol{u}$) → (${}^{S/D}\boldsymbol{\delta}$ ± ${}^{S/D}\boldsymbol{\varepsilon}$) should be desirable provided it ensures (the output-uncertainty) ${}^{S/D}\boldsymbol{\varepsilon}$ as $\leq {}^{S/W}\boldsymbol{u}$. We here, therefore, analyze the implications of **δ**-scale-conversion from the basic standpoint.

## 2. FORMALISM

Generally, even measurement-procedure is (by the aid of relevant standards) established beforehand, i.e. uncertainty ${}^{S/W}\boldsymbol{u}$ is ensured to be acceptably small. However the conversion "(${}^{S/W}\boldsymbol{\delta}$ ± ${}^{S/W}\boldsymbol{u}$) → (${}^{S/D}\boldsymbol{\delta}$ ± ${}^{S/D}\boldsymbol{\varepsilon}$)" should, kike any other evaluation, mean the incorporation of a **desired** systematic **(**i.e. **given** "${}^{S/D}\boldsymbol{\delta}$ vs. ${}^{S/W}\boldsymbol{\delta}$" relationship based**)** change in the measured-data ${}^{S/W}\boldsymbol{\delta}$. Thus, uncertainty ${}^{S/D}\boldsymbol{\varepsilon}$ should[12] also stand for a desired change in the uncertainty ${}^{S/W}\boldsymbol{u}$. For example, if the slope ($\boldsymbol{M_{S/W}^{S/D}}$) of "${}^{S/D}\boldsymbol{\delta}$ vs. ${}^{S/W}\boldsymbol{\delta}$" variation curve should be unity, then the result ${}^{S/D}\boldsymbol{\delta}$ should be equally as comparable as the measured data ${}^{S/W}\boldsymbol{\delta}$. However, for $\boldsymbol{M_{S/W}^{S/D}}$ to be >1 (or, <**1**), ${}^{S/D}\boldsymbol{\delta}$ should be less (or, even **more**) accurate than ${}^{S/W}\boldsymbol{\delta}$. Thus, here, we should really study the behavior of different possible "${}^{S/D}\boldsymbol{\delta}$ vs. ${}^{S/W}\boldsymbol{\delta}$" relationships and so.

However, for simplicity, we henceforth imply that "${}^{S/W}\boldsymbol{\delta} \equiv \boldsymbol{X}$", "${}^{S/D}\boldsymbol{\delta} \equiv \boldsymbol{Y}$"; "${}^{S/W}\boldsymbol{u} \equiv \boldsymbol{u_X}$", and "${}^{S/D}\boldsymbol{\varepsilon} \equiv \boldsymbol{\varepsilon_Y}$". Similarly, true error, though should usually be ever unknown, is in the case of a *measured* estimate "$\boldsymbol{x}$" referred to as $\boldsymbol{\Delta_X}$, and that in the *desired* estimate "$\boldsymbol{y}$" as $\boldsymbol{Đ_Y}$. However, only relative errors should be meaningful[13]. Therefore, we define**:** $\boldsymbol{\Delta_X} = \frac{\Delta X}{X} = \frac{x - X}{X}$; and**:** $\boldsymbol{Đ_Y} = \frac{\mathrm{d}Y}{Y} = \frac{y - Y}{Y}$. Further, any result is generally reported as**:** $\boldsymbol{X} = (\boldsymbol{x} \pm \boldsymbol{u_X})$, which clarifies that "$\boldsymbol{u_X}$" should represent the *possible* [and, therefore, the ***maximum***] *value* of the true-error $\boldsymbol{\Delta_X}$, i.e.[12]**:** $\boldsymbol{u_X}$ =

$^{\text{Max}}|\Delta_X|$; and thus: $\boldsymbol{\varepsilon_Y} = {}^{\text{Max}}|\boldsymbol{Đ_Y}|$. We may refer to "$\boldsymbol{u}$" and "$\boldsymbol{\varepsilon}$" as the measurement [or, *input*] and *output* accuracies (or inaccuracies or uncertainties), respectively[12]. Moreover, considering, e.g.: $\boldsymbol{Y = g(\{X_i\}_{i=1}^{N})}$; i.e. for: $\boldsymbol{(y + Đ_Y) = g(\{x_i + \Delta_i\}_{i=1}^{N})}$; and: $\boldsymbol{(y \pm \varepsilon_Y) = g(\{x_i \pm u_i\}_{i=1}^{N})}$; the output-error ($\boldsymbol{Đ_Y}$) could be shown to be decided as[12,13]:

$$Đ_Y = \sum_{i=1}^{N} (M_i^Y \times \Delta_i) \tag{1}$$

And, the output-uncertainty ($\boldsymbol{\varepsilon_Y}$) could, even a priori, be ascertained as[12,14]:

$$\varepsilon_Y = \sum_{i=1}^{N} (|M_i^Y| \times u_i) = [\sum_{i=1}^{N} (|M_i^Y| \times F_i)] \times {}^{G}u = ([UF]_Y \times {}^{G}u) \tag{2}$$

where $\boldsymbol{M_i^Y}$ is a theoretical constant, representing the relationship "$\boldsymbol{g}$" specific (relative) rate of variation of $\boldsymbol{Y}$ as a function of $\boldsymbol{X_i}$:

$$M_i^Y = \left(\frac{\partial Y}{\partial X_i}\right)\left(\frac{X_i}{Y}\right) = \left(\frac{\partial Y/Y}{\partial X_i/X_i}\right), \quad i = 1, 2 \ldots N \tag{3}$$

And, $\boldsymbol{{}^{G}u}$ is any $\boldsymbol{u_i}$*-value* [viz. which could be believed to be achieved before establishing $\boldsymbol{u_i(s)}$]; $\boldsymbol{F_i} = (\boldsymbol{u_i} / \boldsymbol{{}^{G}u})$; and $\boldsymbol{[UF]_Y}$ may be called[14] as uncertainty-factor. Clearly, if all $\boldsymbol{X_i}$-measurements should be subject to equal uncertainty ($\boldsymbol{u_i} = \boldsymbol{{}^{G}u}$, i.e. if: $\boldsymbol{F_i} = 1$, with: $\boldsymbol{i} = 1, 2 \ldots N$), then:

$$[UF]_Y = \sum_{i=1}^{N} (|M_i^Y| \times F_i) = \sum_{i=1}^{N} |\boldsymbol{M_i^Y}| \tag{4}$$

However the important point is that, if only "$\boldsymbol{[UF]_Y}$ is **<1**", the function as "$\boldsymbol{g}$" should really represent a desirable evaluation method.

Further, let us represent "$\boldsymbol{X \xrightarrow{Ai(s)} Y}$" as: $\boldsymbol{Y = f(X, Z1, Z2 \ldots)}$, cf. below. Then, corresponding $\boldsymbol{\{M_i^Y\}}$ may be referred to as "$\boldsymbol{M_X^Y, M_{Z1}^Y, M_{Z2}^Y \ldots}$"; and $\boldsymbol{\{u_i\}}$ as "$\boldsymbol{u_X, u_{Z1}, u_{Z2} \ldots}$".

We now look into the exact functions as "$\boldsymbol{g}$" (evaluation-methods). As "$\boldsymbol{R}$" cannot be zero or infinity, the expression "$\boldsymbol{Y = ([{}^{S}R/{}^{D}R] - 1)}$" can itself be transformed into the required *scale-conversion* ("$\boldsymbol{{}^{S/W}\delta \rightarrow {}^{S/D}\delta}$" ≡ "$\boldsymbol{X \rightarrow Y}$") formula:

$$Y = \left( \frac{{}^{S}R}{{}^{D}R} - 1 \right) = \left( \frac{{}^{S}R}{{}^{W}R} \times \frac{{}^{W}R}{{}^{D}R} - 1 \right) = \left( [X+1] \times [C+1] - 1 \right) = f(X) \tag{5}$$

where $\boldsymbol{C}$ is the $\boldsymbol{W}$ vs. $\boldsymbol{D}$ isotopic calibration constant ($\boldsymbol{C} = [{}^{W}\boldsymbol{R}/{}^{D}\boldsymbol{R}] - 1$), and should be known.

However, the determining of "$\boldsymbol{C}$" could be avoided by employing[10,15-18] one or more calibrated-***auxiliary***-standards (***A*1**, ***A*2**, etc) in the process**:** $\boldsymbol{X} \rightarrow \boldsymbol{Y}$. The corresponding scale conversion formulae can also, like Eq. 5, be easily derived:

$$Y = \left( \frac{{}^{S}R}{{}^{D}R} - 1 \right) = \left( \frac{{}^{S}R}{{}^{W}R} \times \frac{{}^{W}R}{{}^{A1}R} \times \frac{{}^{A1}R}{{}^{D}R} \right) - 1 = \left( \frac{(X+1) \times (C1+1)}{(Z1+1)} - 1 \right) = f(X, Z1) \tag{6}$$

$$Y = \left( \frac{{}^{S}R}{{}^{W}R} \times \frac{{}^{W}R}{{}^{A1}R - {}^{A2}R} \times \frac{{}^{A1}R - {}^{A2}R}{{}^{D}R} \right) - 1 = \left( (X+1) \times \frac{(C1 - C2)}{(Z1 - Z2)} - 1 \right) = f(X, Z1, Z2) \tag{7}$$

**…**

where ***C*1** and ***C*2** are known (***Ai*** vs. ***D*** calibration) constants [$\boldsymbol{C1} = ([{}^{A1}\boldsymbol{R}/{}^{D}\boldsymbol{R}] - 1)$ and**:** $\boldsymbol{C2} = ([{}^{A2}\boldsymbol{R}/{}^{D}\boldsymbol{R}] - 1)$]; and ***Z*1** and ***Z*2** are the ***Ai***-*measured* variables**:** $\boldsymbol{Z1} = ([{}^{A1}\boldsymbol{R}/{}^{W}\boldsymbol{R}] - 1)$ and**:** $\boldsymbol{Z2} = ([{}^{A2}\boldsymbol{R}/{}^{W}\boldsymbol{R}] - 1)$.

It may, however, be pinpointed that the formula normally used in practice for employing (single ***Ai*** is the same as Eq. 6, but) *two* different ***Ai***-standards is (*different* from Eq. 7) [16,17]:

$$Y = (X - Z2) \times \left( \frac{C1 - C2}{Z1 - Z2} \right) + C2 \tag{8}$$

Unfortunately, Eq. 8 cannot represent a scale conversion method[19]**:** as ***C*1** and ***C*2** are *constants*, and ***Z*1** and ***Z*2** are *variables* [i.e. as**:** $(\boldsymbol{C1} \times \boldsymbol{Z2}) \neq (\boldsymbol{C2} \times \boldsymbol{Z1})$]; the right hand side cannot even be reduced to "$\boldsymbol{Y}$".

However (irrespective how "$\boldsymbol{Y}$" is estimated)**:** $\boldsymbol{Y} = ([{}^{S}\boldsymbol{R}/{}^{D}\boldsymbol{R}] - 1)$; and ${}^{D}\boldsymbol{R}$ is ever known. Therefore, the absolute ratio (${}^{S}\boldsymbol{R}$) can, at least, be readily estimated:

$$ {}^{S}R = {}^{D}R \times (Y + 1) \tag{9}$$

## 3. RESULTS AND DISCUSSION

### 3.1 Appropriate scale conversion method

We, for visualizing a priori which design of evaluation should be appropriate, consider both $\boldsymbol{X}$ and $\boldsymbol{Y}$ as known, viz. sample $\boldsymbol{S}$ and lab-reference $\boldsymbol{W}$ to be (the $^{2}H/^{1}H$ certified materials as) IAEA-CH-7 and GISP, respectively. That is, say that[20]**:** "$\boldsymbol{Y} \equiv {}^{S/D}\boldsymbol{\delta}_{2/1}$" = –0.10033; and "$\boldsymbol{C} \equiv {}^{W/D}\boldsymbol{\delta}_{2/1}$" **=** –0.18973 **(**with $\boldsymbol{D}$ as VSMOW [${}^{D/D}\boldsymbol{\delta}_{2/1}$ = 0, but**:** ${}^{D}\boldsymbol{R}$ = 15.576×10$^{-5}$]**)**; so that**:** ${}^{S}\boldsymbol{R}$ = 14.013260×10$^{-5}$; and ${}^{W}\boldsymbol{R}$ = 12.62076552×10$^{-5}$, and/ or that**:** $\boldsymbol{X} = \boldsymbol{([{}^{S}R/{}^{W}R] - 1)}$ = 0.1103336. Further [cf. Eq.**7**], we choose $\boldsymbol{A1}$ to be NBS-1 ("$\boldsymbol{C1} \equiv {}^{A1/D}\boldsymbol{\delta}_{2/1}$" **=** –0.0476**)**, and $\boldsymbol{A2}$ to be NBS-1A ("$\boldsymbol{C2} \equiv {}^{A2/D}\boldsymbol{\delta}_{2/1}$" **=** –0.1833),[21] i.e.**:** $\boldsymbol{Z1} = (\boldsymbol{[{}^{A1}R/{}^{W}R] - 1})$ = 0.175410665, and**:** $\boldsymbol{Z2} = (\boldsymbol{[{}^{A2}R/{}^{W}R] - 1})$ = 7.9356264×10$^{-3}$. However, for the method as Eq. 6, we use either NBS-1 or NBS-1A as $\boldsymbol{A1}$.

The natures of variations to be expected in the desired **δ-**estimate "$\boldsymbol{y}$" as a function of scale-conversion-method (cf. Eqs. 5-7), and thus in the absolute result "${}^{S}\boldsymbol{r}$" (cf. Eq. 9), are exemplified in Table 1, where *all* method specific *measured estimates* are considered to be ***either*** 100% accurate (cf. example no. 0); ***or*** at ±1% errors (i.e.**:** $\boldsymbol{x = [X \pm 0.01X]}$; and**:** $\boldsymbol{zi = [Zi \pm 0.01Zi]}$; cf. example nos. 1 and 2). It may be noted that, corresponding to example no. 0 (i.e. for**:** $\boldsymbol{x} = \boldsymbol{X}$; and if appropriate**:** $\boldsymbol{zi} = \boldsymbol{Zi}$), the results have also turned out 100% accurate ($\boldsymbol{y} = \boldsymbol{Y}$, and**:** ${}^{S}\boldsymbol{r} = {}^{S}\boldsymbol{R}$; irrespective of method/ BLOCK); thereby clarifying, from purely experimental viewpoint, that any of Eqs.. **5-7** should represent a valid scale conversion method. However, as indicated by example no. 1 and/ or no. 2, Eq. **5** should yield the most accurate $\boldsymbol{Y}$-value. In other words, the employing of $\boldsymbol{Ai}$-standard(s), viz. as Eq. **6**/ Eq. **7**, should cause the desired **δ**-estimate **(**$\boldsymbol{y}$, and hence the absolute estimate ${}^{S}\boldsymbol{r}$) to be**:** (i) rather **inaccurate**, and**:** (ii) ***Ai(s)****-specific* [comparison between BLOCK Nos. **2** and **2a**]. Further, it should be interesting to note that, irrespective of method, "$^{S}\boldsymbol{r}$" is reflected to be *less* erroneous than "***y***" by a constant-factor as "$|\mathbf{Đ}_{Y}|/|\mathbf{Đ}_{R}|$".

### 3.2 Can the (accuracy of) desired $^{S/D}\delta$-estimate be decided by the lab-reference *W*?

In order for ascertaining whether scale-conversion can make the estimate "***y***" to be free of the measurement-reference ***W***, we now replace the ***W***-material, GISP, by e.g. SLAP **(**i.e. consider**:** "$\boldsymbol{C} \equiv {}^{W/D}\boldsymbol{\delta}_{2/1}$" **=** –0.428,[20] and hence**:** $^{W}\boldsymbol{R}$ = 8.909472×10$^{-5}$, and/ or**:** ***X*** = 0.57284965, ***Z*1** = 0.665034965, and ***Z*2** = 0.427797203**)**, reevaluate all the results (as Table 1) and present them in Table 2. However, the new estimates as example no. 1 or 2 [cf. *any* BLOCK in Table 2] are ***different*** from (truly, more *erroneous* than) the corresponding estimates in Table 1. That is, even the **δ**-estimate as ***y***, and thus the absolute estimate $^{S}\boldsymbol{r}$ (and/ or achievable accuracies, $\boldsymbol{\varepsilon}_{Y}$ and $^{S}\boldsymbol{\varepsilon}_{R}$, respectively), are signified to be ***W***-*specific*. Moreover, the indication (Table 1) that "$^{S/D}\boldsymbol{\delta} \equiv \boldsymbol{y}$" to "$^{S}\boldsymbol{r}$" conversion helps improve accuracy is rather confirmed in Table 2, which signifies the accuracy-enhancement-factor "$|\mathbf{Đ}_{Y}|/|\mathbf{Đ}_{R}|$" to be independent of *even* "***W***".

### 3.3 Are the above findings fictitious?

The uncertainties $\boldsymbol{\varepsilon}_{Y}$ and $^{S}\boldsymbol{\varepsilon}_{R}$ [with**:** $^{G}\boldsymbol{u}$ = 1%], i.e. the uncertainty-factors $\boldsymbol{[UF]_{Y}}$ and $\boldsymbol{[UF]_{R}}$ (for determining "***Y***" [by either of the methods as Eqs. 5-7] and the [corresponding] absolute-ratio $^{S}\boldsymbol{R}$, respectively), are exemplified, and even in terms of their governing factors illustrated, in Table 3, where any **1$^{st}$ set** of data (e.g. "$\boldsymbol{M_{X}^{Y}}$ = –0.891") relates to the employing of GISP as "***W***" (cf. Table 1), and any 2$^{nd}$ set of data [e.g.**:** $\boldsymbol{M_{X}^{Y}}$ = "**(–3.266)**"] corresponds to SLAP as "***W***" [cf. Table 2]. In any case, the ***W***-specific predictions [for ***W*** as GISP and, e.g. scale conversion method as Eq. 5]**:** $\boldsymbol{\varepsilon}_{Y}^{\mathrm{Eq.5}}$ = ($\boldsymbol{[UF]_{Y}}$ × $^{G}\boldsymbol{u}$) = (0.891 × $^{G}\boldsymbol{u}$) = 0.891**%**; and [even for ***W*** to be SLAP]**:** $\boldsymbol{\varepsilon}_{Y}^{\mathrm{Eq.5}}$ = 3.266**%**; are in corroboration with the findings as **(**$|\mathbf{Đ}_{Y}|$ = 0.891**%** and**:** $|\mathbf{Đ}_{Y}|$ = 3.266**%** in**)** Tables 1 and 2, respectively.

3.3.1 *Why should the results be lab-reference* (***W***) *specific*?

Perhaps, variation of (the IC of the) lab-reference ***W*** is believed to affect the achievable-measurement-accuracy $^{G}\boldsymbol{u}$ only. That is the estimates **(**e.g. corresponding to Eq. 5 and example no. 1, but which were obtained for ***equal*** [1%] *errors* in the ***W***-specific $^{SW}\boldsymbol{\delta}$-estimates "$\boldsymbol{x}^{\mathbf{(GISP)}}$ and $\boldsymbol{x}^{\mathbf{(SLAP)}}$"**)** were not expected to vary between Tables 1 and 2. —— However, it is already illustrated by the considerations as Eqs 1-3 above that the "$\boldsymbol{x} \rightarrow \boldsymbol{y}$" translation should be equivalent to the error-transformation as "$\boldsymbol{\Delta_X} \xrightarrow{M_X^Y} \boldsymbol{Đ_Y}$", or "$\boldsymbol{u_X} \xrightarrow{|M_X^Y|} \boldsymbol{\varepsilon_Y}$". Further (cf. Table 3)**:** $\boldsymbol{M_X^Y}$ = [($^{S}\boldsymbol{R}$ – $^{W}\boldsymbol{R}$**)** / **(**$^{S}\boldsymbol{R}$ – $^{D}\boldsymbol{R}$**)**]; i.e. the $^{S/D}\boldsymbol{\delta}$-estimate, ***y***, has to be ***W***-specific. Again, "$^{SLAP}\boldsymbol{R}$" is "<$^{GISP}\boldsymbol{R}$", which explains why the results for employing SLAP (cf. Table **2**), than GISP (cf. Table **1**), as ***W*** are more *erroneous*.

3.3.2 *Why could the results vary for simply the choice of* ***Ai****-standard(s)*?

Table 3 (cf. for Eq. 6) confirms that the employing of even single ***auxiliary***-reference ***A*1** should cause "***y***" to be (***Ai***-specific and/ or) subject to additional measurement-variation at the rate as**:** $\boldsymbol{M_{Z1}^Y}$ = (**[**$^{S}\boldsymbol{R}$**(**$^{W}\boldsymbol{R}$ – $^{A1}\boldsymbol{R}$**)]/[**$^{A1}\boldsymbol{R}$**(**$^{S}\boldsymbol{R}$ – $^{D}\boldsymbol{R}$**)]**). Further, "|$^{W}\boldsymbol{R}$ – $^{NBS\text{-}1A}\boldsymbol{R}$|" is "< |$^{W}\boldsymbol{R}$ – $^{NBS\text{-}1}\boldsymbol{R}$|" (with ***W*** as either GISP or SLAP). This explains why, even though the ***A*1**-measurement-accuracy $\boldsymbol{u_{Z1}}$ is considered to be unchanged, the employing of ***A*1** as NBS-1A (cf. block **2a** in Table **1** or **2**), rather than as NBS-1 (cf. block **2**), is observed to yield *better* representative result(s).

3.3.3 *Should the aid of* ***Ai****-standards be at all worth*?

It was often meant in the literature[10,16-18] that the method as Eq. 7 ($\boldsymbol{X} \xrightarrow{A1,A2} \boldsymbol{Y}$), specifically for employing sample bracketing ***Ai***-standards, should yield more accurate estimate "***y***" than the method as Eq. 6 ($\boldsymbol{X} \xrightarrow{A1} \boldsymbol{Y}$) or Eq. 5 ($\boldsymbol{X} \rightarrow \boldsymbol{Y}$). Unfortunately, the selection of sample-bracketing

***Ai***-references should require the desired-unknown "***Y***" itself to be known beforehand. Again, as shown by considering a *known* case ("$^{S/D}\boldsymbol{\delta} \equiv \boldsymbol{Y}$" = –0.10033) here, "***y***" obtained by employing Eq. 7 **(**with sample-bracketing ***Ai***-standards ["$^{A1/D}\boldsymbol{\delta} \equiv {}^{\text{NBS-1}/D}\boldsymbol{\delta}$" = –0.0476; and "$^{A2/D}\boldsymbol{\delta} \equiv {}^{\text{NBS-1A}/D}\boldsymbol{\delta}$" **=** –0.1833]; cf. example no. 1/ 2 in BLOCK no. 3 of Table 1 or 2**)** is more **erroneous** than the "***y***" obtained for using *either* NBS-1 (cf. BLOCK no. 2) *or* NBS-1A (cf. BLOCK no. 2a] as ***A*1** *or* for no ***Ai*** (cf. BLOCK no. 1). In any case, the rate-of-variation ($\boldsymbol{M_X^Y}$) of ***Y*** as a function of ***X*** (i.e. the effect of any possible error, $\boldsymbol{u_X}$, in measuring the sample ***S***, on the desired result "***y***") is shown to be ***fixed***, *irrespective* scale-conversion-method (cf. Table 3). Thus, "***y***" by Eq. **6/ 7** can never be more accurate than the "***y***" by Eq. **5**.

Further, Eq. 6 requires only one ***Ai***-*measurement*, but Eq. 7 involves two. That is, uncertainty "$\boldsymbol{\varepsilon_Y}^{\text{Eq.7}}$" should even be expected to be "$>\boldsymbol{\varepsilon_Y}^{\text{Eq.6}}$". Yet, it may be pointed out that "$\boldsymbol{M_{Z1}^Y}$" (and hence "$\boldsymbol{\varepsilon_Y}^{\text{Eq.6}}$", cf. Table 3 for Eq. 6) could be varied by varying either or both "***W***- and ***A*1**-materials" [i.e. the difference "$|{}^{W}\boldsymbol{R} - {}^{A1}\boldsymbol{R}|$" and the ratio "$^{S}\boldsymbol{R}/{}^{A1}\boldsymbol{R}$"]; but "$\boldsymbol{\varepsilon_Y}^{\text{Eq.7}}$" (i.e. "$\boldsymbol{M_{Z1}^Y}$" and "$\boldsymbol{M_{Z2}^Y}$" corresponding to Eq. 7) should vary as a function of even the difference "$|{}^{A1}\boldsymbol{R} - {}^{A2}\boldsymbol{R}|$". Thus, though "$\boldsymbol{\varepsilon_Y}^{\text{Eq.7}}$" cannot (in any unknown case) be ensured to be "$<\boldsymbol{\varepsilon_Y}^{\text{Eq.6}}$", *hypothetical-systems* [of the type as**:** "$\mathbf{1} < ({}^{S}\boldsymbol{R}/{}^{A1}\boldsymbol{R}) > ({}^{S}\boldsymbol{R}/{}^{A2}\boldsymbol{R})$"] may be designed to yield**:** $\boldsymbol{\varepsilon_Y}^{\text{Eq.7}} < \boldsymbol{\varepsilon_Y}^{\text{Eq.6}}$.

### 3.3.4 *Why did absolute estimate* ($^{S}\boldsymbol{r}$) *turn out more accurate than* $^{S/D}\boldsymbol{\delta}$*-estimate* (***y***)?

The estimate "$^{S}\boldsymbol{r}$" is obtained from the estimate "***y***" (cf. Eq. 9), and is why (cf. Table 3)[18,19]**:** $^{S}\boldsymbol{\varepsilon_R} = (|\boldsymbol{M_Y^R}| \times \boldsymbol{\varepsilon_Y}) = (|\boldsymbol{M_Y^R}| \times \boldsymbol{[UF]_Y} \times {}^{G}\boldsymbol{u}) = (\boldsymbol{[UF]_R} \times {}^{G}\boldsymbol{u})$. Further**:** $\boldsymbol{M_Y^R} = ([{}^{S}\boldsymbol{R} - {}^{D}\boldsymbol{R}]/{}^{S}\boldsymbol{R})$, i.e. "$|\boldsymbol{M_Y^R}|$" has to be **<1**, and/ or *error-reduction* ["$({}^{S}\boldsymbol{\varepsilon_R}/\boldsymbol{\varepsilon_Y}) \equiv (\boldsymbol{[UF]_R}/\boldsymbol{[UF]_Y})$" < **1**] should be a feature of the transformation "$\boldsymbol{y} \rightarrow {}^{S}\boldsymbol{r}$". Over and above, if only sample ***S*** and standard ***D*** are fixed, then "$\boldsymbol{M_Y^R}$" should also be fixed. This explains why the *accuracy-enhancement-factor*

"$(|\boldsymbol{Đ}_{Y}|/|\boldsymbol{Đ}_{R}|) \equiv [\varepsilon_{Y}/{}^{S}\varepsilon_{R}]$" is observed (cf. Tables 1 and 2) to be fixed (as "$[\mathbf{1}/|M_{Y}^{R}|] = 8.9671$") and/ or independent of the measurement-reference ***W***, scale-conversion-method, and ***Ai***-material(s).

### 3.4 Could our findings be trivial (source-specific)?

Remembering that [output-uncertainty, cf. Eq. 2]**:** $\boldsymbol{\varepsilon} = ([UF] \times {}^{G}u)$, the variations of $[UF]_{Y}$ and $[UF]_{R}$ **(**i.e. uncertainty-factors for **δ**-scale-conversion by each of the methods as Eqs. 5-7, and for determining [**δ**-method-specific value of] the sample-isotopic-ratio ${}^{S}R$ [cf. Eq. 9]**)** as a function of the measurable ${}^{S/W}\boldsymbol{\delta}$-quantity "***X***" **(**with a specific material [GISP] as "***W***", and hence against ${}^{S}R$ itself, cf. the *top-axis***)**, are depicted in Figures 1 and 2, respectively. Clearly (cf. the considerations for innumerable ***S***-sources as the IAEA-CH-7 "$X = 0.1103336$" in. Fig. **1**), the above finding "$\varepsilon_{Y}^{Eq.7} > \varepsilon_{Y}^{Eq.6} > \varepsilon_{Y}^{Eq.5}$" has to be a *source-independent* fact.

#### 3.4.1 *Can any* ${}^{S/D}\boldsymbol{\delta}$*-estimate "**y**" be free of **W***?

As Fig. 1 implies, "$[UF]_{Y}^{Eq.5}$" **(**actually, as already clarified in Table 3, the rate-of-variation "$|M_{X}^{Y}|$", i.e. $[UF]_{Y}$ but relating to ***S****-measurement* alone, and hence which is an integral part of *any method-specific-*$[UF]_{Y}$**)** should decrease for decreasing "$|X|$" (rather "$|{}^{S}R - {}^{W}R|$" only), and be *zero* (if ***S*** and ***W*** happen to have the same IC, i.e.) at "$X = 0$". Thus, the other observations, viz.**:** (i) **δ**-scale-conversion doesn't make the estimate "***y***" free from the measurement-reference ***W***, but (ii) accuracy of any method-specific "***y***" is improved for simply considering ***W*** to be closer to the sample-***S*** [in IC], should also represent general facts. Above all, the prediction that the measurement of even the ${}^{Ai/W}\boldsymbol{\delta}$-quantity (***Zi***) should cause ***y*** to be ***W***-specific (cf. Table 3, which clarifies the rate-of-variation "$M_{Zi}^{Y}$" to also be decided by, among others, "${}^{W}R$") is verified by the fact that neither "$[UF]_{Y}^{Eq.6}$" nor "$[UF]_{Y}^{Eq.7}$" is zero at "$X = 0$" (cf. Fig. 1, which has considered the ***Ai(s)*** as being different [in IC] from ***W***).

3.4.2 *Variation-identifying tool*: ${}^{S/D}\boldsymbol{\delta}$*-estimate* (***y***) *or absolute-estimate* (${}^{S}\boldsymbol{r}$)?

Usually, any measured (e.g. ${}^{S/W}\boldsymbol{\delta}$) estimate (***x***) is a priori ensured to be accurate. Similarly, (corresponding) reference-standard ***D*** is also a prefixed one. Thus, sample-***S*** can happen to be very close, in IC, to ***D***. However, what is revealed here (cf. Fig. 1 for "***X*** = 0.235" or so) is that the ${}^{S/D}\boldsymbol{\delta}$-estimate ***y*** should, for "$|{}^{S}\boldsymbol{R} - {}^{D}\boldsymbol{R}| \to \mathbf{0}$", turn out *increasingly* ***erroneous***.

Nevertheless, for ***D*** to be very different (viz. here, a highly $^{2}$**H**-enriched material) from any corresponding unknown-***S***, "***y***" would be accurate (cf. Fig. 1 for: ***X*** = ±0.6 or so). Unfortunately, the latter should cause the *measure-of-variation* "$|{}^{S}\boldsymbol{R} - {}^{D}\boldsymbol{R}|$" itself to be ***larger*** *than* the source-value "${}^{S}\boldsymbol{R}$" (i.e. [cf. Table 3] "$|M_Y^R|$" to be **>1)**, and hence be inconceivable as a viable proposal.

In any case (cf. Table 3), the product "$|M_Y^R|$ × ***[UF]***$_Y$" defines "***[UF]***$_R$". This is why, the variation of the *method-independent* parameter "$|M_Y^R|$" (against "***X***") is also described as an insert in Fig. 1, thereby explaining why (although "***[UF]***$_Y$-curves" pass through the **peak** as infinity) "***[UF]***$_R$-curves" [cf. Fig. **2**] are of **valley**-shapes. However (cf. here above), any real world "$|M_Y^R|$" should be **<1**, i.e. the above finding "uncertainty-${}^{S}\boldsymbol{\varepsilon}_R$ is <$\boldsymbol{\varepsilon}_Y$" has also to be a general one. In other words, "$|M_Y^R|$ > **1** (cf. Fig. 1)", and hence "${}^{S}\boldsymbol{\varepsilon}_R$ > $\boldsymbol{\varepsilon}_Y$", should represent hypothetical cases. Moreover, for unknown-***S*** to be close to ***D***, the ${}^{S/D}\boldsymbol{\delta}$-estimate ***y*** may even misrepresent "***S***" (cf. Fig. 1**: *[UF]***$_Y$ ≫ **1**). However, what is interesting (cf. Fig. **2**, e.g. the ***[UF]***$_R^{\text{Eqs.(5\&9)}}$-curve) is that any real world absolute estimate "${}^{S}\boldsymbol{r}$" (specifically, for ${}^{S}\boldsymbol{R} \approx {}^{D}\boldsymbol{R}$, i.e. corresponding to a highly ***inaccurate*** ${}^{S/D}\boldsymbol{\delta}$-estimate "***y***") should be (highly) ***accurate***.

Further, for a case characterized by either "***[UF]***$_Y$ **= 0**" or "$|M_Y^R|$ **= 0**", ***[UF]***$_R$ (and hence ${}^{S}\boldsymbol{\varepsilon}_R$) should also equal *zero*. Thus, as Fig. 2 clarifies, any method-specific "*minima*" corresponds to "***X*** **= 0** (i.e. ***S*** as identical with ***W***)". For example, "${}^{Min.}$***UF]***$_R^{\text{Eqs.(5\&9)}}$" equals *zero*, and is because that (cf. Fig. 1): $^{Min.}\boldsymbol{UF]_Y}^{\text{Eq.5}}$ = **0**. However, why <u>shouldn't</u> "$\left|M_Y^R\right|$ = **0**" (<u>***S*** as identical with ***D***</u>, cf. the insert in Fig. 1 for: ***X*** = 0.2341565) cause even corresponding-$\boldsymbol{UF]_R}^{\text{Eqs.(5\&9)}}$ to be <u>zero</u>?

As "$|^S\boldsymbol{R} - {}^D\boldsymbol{R}| \rightarrow \mathbf{0}$" simultaneously implies "$\left|M_X^Y\right| \rightarrow \infty$" (i.e.: $\boldsymbol{[UF]_Y} \rightarrow \infty$) and "$\left|M_Y^R\right| \rightarrow \mathbf{0}$"; the corresponding $\boldsymbol{[UF]_R}$ should be decided (cf. Table 3: $\boldsymbol{[UF]_R}^{\text{Eqs.(5\&9)}} = \left|\mathbf{1} - ({}^W\boldsymbol{R}/{}^S\boldsymbol{R})\right|$; and hence) as: $^{Lim.(S\rightarrow D)}\boldsymbol{[UF]_R}^{\text{Eqs.(5\&9)}} = \left|\mathbf{1} - ({}^W\boldsymbol{R}/{}^D\boldsymbol{R})\right| = \left|\boldsymbol{C}\right|$ = 0.18973. Similarly, one may verify that: $^{Lim.(S\rightarrow D)}\boldsymbol{[UF]_R}^{\text{Eqs.(6\&9)}} = (\left|\boldsymbol{C}\right| + \left|({}^W\boldsymbol{R} - {}^{A1}\boldsymbol{R})/{}^{A1}\boldsymbol{R}\right|) = (\left|\boldsymbol{C}\right| + \left|(\boldsymbol{C} - \boldsymbol{C}\mathbf{1})/(\boldsymbol{C}\mathbf{1} + \mathbf{1})\right|)$ = 0.33896; and: $^{Lim.(S\rightarrow D)}\boldsymbol{[UF]_R}^{\text{Eqs.(7\&9)}} = (\left|\boldsymbol{C}\right| + (\left|{}^W\boldsymbol{R} - {}^{A1}\boldsymbol{R}\right| + \left|{}^{A2}\boldsymbol{R} - {}^W\boldsymbol{R}\right|)/\left|{}^{A1}\boldsymbol{R} - {}^{A2}\boldsymbol{R}\right|)$

$= (\left|\boldsymbol{C}\right| + (\left|\boldsymbol{C} - \boldsymbol{C}\mathbf{1}\right| + \left|\boldsymbol{C}\mathbf{2} - \boldsymbol{C}\right|)/\left|\boldsymbol{C}\mathbf{1} - \boldsymbol{C}\mathbf{2}\right|)$ = 1.28450.

We may, for illustration, consider a specific unknown-source (***uS)*** to be "$^{uS}\boldsymbol{R}$ = 15.6×10$^{-5}$ (i.e.: $^u\boldsymbol{X} = ([{}^{uS}\boldsymbol{R}/{}^W\boldsymbol{R}] - 1) = ([{}^{uS}\boldsymbol{R}/{}^{GISP}\boldsymbol{R}] - 1)$ = 0.23605814 and: $^u\boldsymbol{Y} = ([{}^{uS}\boldsymbol{R}/{}^D\boldsymbol{R}] - 1)$ = 1.5408291×10$^{-3}$)". Then, as Fig. 1 predicts: $^u\boldsymbol{\varepsilon_Y}^{\text{Eq.5}} = ({}^u\boldsymbol{[UF]_Y}^{\text{Eq.5}} \times {}^G\boldsymbol{u}) = 124.1{}^G\boldsymbol{u}$; and as Fig. 2 implies: $^{uS}\boldsymbol{\varepsilon_R}^{\text{Eqs.(5\&9)}} = [(\left|M_Y^R\right| \times {}^u\boldsymbol{\varepsilon_Y}^{\text{Eq.5}}) = (\left|M_Y^R\right| \times {}^u\boldsymbol{[UF]_Y}^{\text{Eq.5}} \times {}^G\boldsymbol{u}) =] ({}^u\boldsymbol{[UF]_R}^{\text{Eqs.(5\&9)}} \times {}^G\boldsymbol{u})$ = $0.191{}^G\boldsymbol{u}$. That is, for a possible measurement-error $^G\boldsymbol{u}$, the absolute-estimate ($^{uS}\boldsymbol{r}$) should turn out ≈650 times more *accurate* than the $^{uS/D}\boldsymbol{\delta}$-estimate ($^u\boldsymbol{y}$).

However, say, measurement (for ***uS*** by Lab1) has yielded: $^u\boldsymbol{x}^{\textbf{Lab1}} = ({}^u\boldsymbol{X} + {}^u\Delta_X) = ({}^u\boldsymbol{X}$ + 0.1%) = 0.236294. That is, one may verify (cf. Eq. 5): $^u\boldsymbol{y}^{\textbf{Lab1}}$ = 0.0017321 = $({}^u\boldsymbol{Y} + {}^u\boldsymbol{Đ}_Y) = ({}^u\boldsymbol{Y}$ + 12.4%); and in turn (cf. Eq. 9): $^{uS}\boldsymbol{r}^{\textbf{Lab1}}$ = 15.60298×10$^{-5}$ = $({}^{uS}\boldsymbol{R} + {}^{uS}\boldsymbol{Đ}_R) = ({}^{uS}\boldsymbol{R}$ + 0.0191%). Clearly, the results are in corroboration of the above predictions. However, the point to note is that the **true** variation in ***uS*** *from* ***D*** is **(**$[{}^{uS}\boldsymbol{R} - {}^D\boldsymbol{R}]$ = 0.024×10$^{-5}$ = **0.1508291**%, which is**)** more or less accurately represented by the *estimate* $^{uS}\boldsymbol{r}^{\textbf{Lab1}}$ (as: $[{}^{uS}\boldsymbol{r}^{\textbf{Lab1}} - {}^D\boldsymbol{R}]$ = 0.027×10$^{-5}$ = **0.173**%), and rather *misrepresented* by the $^{uS/D}\boldsymbol{\delta}$*-estimate* $^u\boldsymbol{y}^{\textbf{Lab1}}$ (as: $[{}^u\boldsymbol{y}^{\textbf{Lab1}} - {}^u\boldsymbol{Y}]$ = 1.913×10$^{-4}$ = = **12.4**%).

Similarly, another equally suitable lab might be considered to yield**:** ${}^{u}\boldsymbol{x}^{\mathbf{Lab2}}$ = (${}^{u}\boldsymbol{X}$ − 0.1%) = 0.235822, thereby giving "${}^{u}\boldsymbol{y}^{\mathbf{Lab2}}$ = 0.001350 = (${}^{u}\boldsymbol{Y}$ − 12.4%)" and "${}^{uS}\boldsymbol{r}^{\mathbf{Lab2}}$ = 15.59702×10$^{-5}$ = (${}^{uS}\boldsymbol{R}$ − 0.0191%)". Thus the *scatter*, while between the ***absolute*** lab-results is only **0.027**%, between the differential (${}^{uS/D}\boldsymbol{\delta}$) lab-results is as high as **17.5**%.

## 4. CONCLUSIONS

The above study clarifies that different absolute lab-results (${}^{S}\boldsymbol{r}^{\mathbf{Lab1}}$, ${}^{S}\boldsymbol{r}^{\mathbf{Lab2}}$ **…**) should always be more closely comparable than their desired-relative (i.e. ${}^{S/D}\boldsymbol{\delta}$) estimates ($\boldsymbol{y}^{\mathbf{Lab1}}$, $\boldsymbol{y}^{\mathbf{Lab2}}$ **…**). In other words, the variation, if at all any, in a source ***S*** (from a source as ***D***, i.e. as a function of time and so) could be accurately ascertained by the corresponding absolute estimate ${}^{S}\boldsymbol{r}$ (viz. as "$|{}^{S}\boldsymbol{r} - {}^{D}\boldsymbol{R}|$") rather than by the ${}^{S/D}\boldsymbol{\delta}$-estimate ***y***. This is, as shown above, best supported by the fact that the ${}^{S/D}\boldsymbol{\delta}$-estimate ***y*** can turn out to even be *non-representative* of source-***S*** (e.g. in a case where ***S*** could be close, in IC, to the reference-standard ***D***). Most importantly, the "${}^{S}\boldsymbol{r}$" (*corresponding to***:** ${}^{S}\boldsymbol{R} \approx {}^{D}\boldsymbol{R}$) should be (*highly*) accurate.

It is demonstrated that, and also explained why**:** **(i)** the scale conversion (as either**:** "${}^{S/W}\boldsymbol{\delta} \rightarrow {}^{S/D}\boldsymbol{\delta}$" ≡ "$\boldsymbol{X} \rightarrow \boldsymbol{Y}$" or**:** "${}^{S/W}\boldsymbol{\delta} \xrightarrow{Ai(s)} {}^{S/D}\boldsymbol{\delta}$") cannot make the results (the ${}^{S/D}\boldsymbol{\delta}$-estimate ***y***, and hence the absolute estimate ${}^{S}\boldsymbol{r}$) free from the measurement-reference ***W***; and**:** **(ii)** the employing of (*increasing* number of) ***Ai***-standards should cause the estimates (***y*** and ${}^{S}\boldsymbol{r}$) to be, though assumed[10,15-18] accurate, (*increasingly*) *inaccurate*, however.

**Table 1.** Examples of variations in the scale-conversion-method specific ${}^{S/D}\boldsymbol{\delta}$-estimate (***y***), and thus in the absolute-estimate (${}^{S}\boldsymbol{r}$), as a function of required (${}^{S/W}\boldsymbol{\delta}$- and, if applicable, ${}^{Ai/W}\boldsymbol{\delta}$-) data of a given accuracy (${}^{G}\boldsymbol{u}$ = 1%)**:** use of GISP as the measurement-reference ***W***

| BLOCK (Method) No. (Eq. No.) | Example No. | Input (measured) ${}^{S/W}\boldsymbol{\delta}$-estimate $x$, its (relative) error $\Delta_X$, ${}^{Ai/W}\boldsymbol{\delta}$-estimate $zi$, and its error $\Delta_{Zi}$ | | | Output (${}^{S/D}\boldsymbol{\delta}$, or ***absolute***) estimate and (relative) error **Đ** | | Error-ratio $\frac{\lvert Đ_Y \rvert}{\lvert Đ_R \rvert}$ | Projected output accuracy **ε** | |
|---|---|---|---|---|---|---|---|---|---|
| | | $x$ <br>($\Delta_X \times 10^2$) | $z1$ <br>($\Delta_{Z1} \times 10^2$) | $z2$ <br>($\Delta_{Z2} \times 10^2$) | $y$ <br>($Đ_Y \times 10^2$) | ${}^{S}r \times 10^5$ <br>($Đ_R \times 10^2$) | | $\varepsilon_Y \times 10^2$ | ${}^{S}\varepsilon_R \times 10^2$ |
| 1 (Eq. 5) | 0 | 0.11033360 (0) | - | - | –0.100330 (0) | 14.013260 (0) | 0 | 0.891 | 0.1 |
| | 1 | 0.11143694 (1.0) | - | - | –0.0994360 (–0.891) | 14.027185 (0.0994) | 8.9671 | | |
| | 2 | 0.10923026 (–1.0) | - | - | –0.1012234 (0.891) | 13.999335 (–0.0994) | 8.9671 | | |
| 2 (Eq. 6, with ***A*1** as NBS-1) | 0 | 0.11033360 (0) | 0.175410665 (0) | - | –0.100330 (0) | 14.013260 (0) | 0 | 2.23 | 0.25 |
| | 1 | 0.11143694 (1.0) | 0.1736566 (-1.0) | - | –0.0980901 (–2.23) | 14.048149 (0.249) | 8.9671 | | |
| | 2 | 0.10923026 (–1.0) | 0.1771648 (1.0) | - | –0.1025633 (2.226) | 13.978474 (–0.2482) | 8.9671 | | |
| 3 (Eq. 7) | 0 | 0.11033360 (0) | 0.175410665 (0) | $7.9356264 \times 10^{-3}$ (0) | –0.100330 (0) | 14.013260 (0) | 0 | 10.8 | 1.21 |
| | 1 | 0.11143694 (1.0) | 0.1736566 (-1.0) | $8.01498 \times 10^{-3}$ (1.0) | –0.089468 (-10.83) | 14.182447 (1.207) | 8.9671 | | |
| | 2 | 0.10923026 (–1.0) | 0.1771648 (1.0) | $7.85627 \times 10^{-3}$ (–1.0) | –0.110957 (10.59) | 13.847732 (–1.181) | 8.9671 | | |
| 2a (Eq. 6, with ***A*1** as NBS-1A) | 0 | 0.11033360 (0) | $7.9356264 \times 10^{-3}$ (0) | - | –0.100330 (0) | 14.013260 (0) | 0 | 0.962 | 0.11 |
| | 1 | 0.11143694 (1.0) | $7.85627 \times 10^{-3}$ (–1.0) | - | –0.0993651 (–0.9617) | 14.028289 (0.1073) | 8.9671 | | |
| | 2 | 0.10923026 (–1.0) | $8.01498 \times 10^{-3}$ (1.0) | - | –0.1012948 (0.9616) | 13.998233 (–0.1072) | 8.9671 | | |

**Table 2.** Variations of scale-conversion-method specific ${}^{S/D}\boldsymbol{\delta}$-estimate (***y***), and thus of absolute-estimate (${}^{S}\boldsymbol{r}$), and/ or (their) accuracies for employing SLAP (i.e. instead of GISP, cf. Table 1) as the measurement-reference ***W***

| Block (Method) No. (Eq. No.) | Example No. | Input (measured) ${}^{S/W}\boldsymbol{\delta}$-estimate ***x***, its (relative) error $\Delta_X$, ${}^{Ai/W}\boldsymbol{\delta}$-estimate ***zi***, and its error $\Delta_{Zi}$ | | | Output (${}^{S/D}\boldsymbol{\delta}$, or ***absolute***) estimate and (relative) error **Đ** | | Error-ratio $\frac{\lvert Đ_Y \rvert}{\lvert Đ_R \rvert}$ | Projected output accuracy **ε** | |
|---|---|---|---|---|---|---|---|---|---|
| | | ***x*** ($\Delta_X \times 10^2$) | ***z1*** ($\Delta_{Z1} \times 10^2$) | ***z2*** ($\Delta_{Z2} \times 10^2$) | ***y*** ($Đ_Y \times 10^2$) | ${}^{S}\boldsymbol{r} \times 10^5$ ($Đ_R \times 10^2$) | | $\varepsilon_Y \times 10^2$ | ${}^{S}\varepsilon_R \times 10^2$ |
| 1<br>(Eq. 5) | 0 | 0.57284965<br>(0) | - | - | –0.100330<br>(0) | 14.013260<br>(0) | 0 | 3.266 | 0.364 |
| | 1 | 0.57858<br>(1.0) | - | - | –0.097053<br>(–3.266) | 14.06430<br>(0.3642) | 8.9671 | | |
| | 2 | 0.56712<br>(–1.0) | - | - | –0.103607<br>(3.266) | 13.962222<br>(–0.3642) | 8.9671 | | |
| 2<br>(Eq. 6,<br>with ***A1*** as<br>NBS-1) | 0 | 0.57284965<br>(0) | 0.665034965<br>(0) | - | –0.100330<br>(0) | 14.013260<br>(0) | 0 | 6.87 | 0.767 |
| | 1 | 0.57858<br>(1.0) | 0.6583846<br>(-1.0) | - | –0.093432<br>(–6.87) | 14.12070<br>(0.7667) | 8.9671 | | |
| | 2 | 0.56712<br>(–1.0) | 0.6716853<br>(1.0) | - | –0.107173<br>(6.82) | 13.906677<br>(–0.7606) | 8.9671 | | |
| 3<br>(Eq. 7) | 0 | 0.57284965<br>(0) | 0.665034965<br>(0) | 0.427797203<br>(0) | –0.100330<br>(0) | 14.013260<br>(0) | 0 | 46.7 | 5.2 |
| | 1 | 0.11143694<br>(1.0) | 0.6583846<br>(-1.0) | 0.432075<br>(1.0) | –0.05345<br>(-46.7) | 14.74345<br>(5.21) | 8.9671 | | |
| | 2 | 0.10923026<br>(–1.0) | 0.6716853<br>(1.0) | 0.423519<br>(–1.0) | –0.14308<br>(42.6) | 13.34738<br>(–4.75) | 8.9671 | | |
| 2a<br>(Eq. 6,<br>with ***A1*** as<br>NBS-1A) | 0 | 0.57284965<br>(0) | 0.427797203<br>(0) | - | –0.100330<br>(0) | 14.013260<br>(0) | 0 | 6.0 | 0.67 |
| | 1 | 0.11143694<br>(1.0) | 0.432075<br>(1.0) | - | –0.094340<br>(–5.97) | 14.106564<br>(0.666) | 8.9671 | | |
| | 2 | 0.10923026<br>(–1.0) | 0.423519<br>(–1.0) | - | –0.106284<br>(5.935) | 13.920513<br>(–0.662) | 8.9671 | | |

**Table 3:** Scale-conversion (**SC**) processes (and also the process as Eq. 9) specific parameters

| SC-Formula | Process-specific rate-of-variation ($M_i^Y$, cf. Eq. 3) | $[UF]_Y$ (cf. Eq. 4) | Output-Uncertainty (ε, cf. Eq. 2) | | $\varepsilon_Y/{}^S\varepsilon_R$ |
|---|---|---|---|---|---|
| | | | $\varepsilon_Y$ | ${}^S\varepsilon_R$ $^{*1}$ | |
| Eq. 5 | $M_X^Y = \left(\frac{dY}{dX}\right)\frac{X}{Y} = \frac{X\,(C+1)}{(X+1)\,(C+1)-1} = \frac{{}^SR-{}^WR}{{}^SR-{}^DR} = -0.891^{*2}; (-3.266)^{*3}$ | $[UF]_Y^{Eq.5} = \lvert M_X^Y \rvert = 0.891^{*2}$; $(3.266)^{*3}$ | $\varepsilon_Y^{Eq.5} = ([UF]_Y^{Eq.5} \times {}^Gu) = 0.891\%^{*2}$ $(3.266\%)^{*3}$ | ${}^S\varepsilon_R^{Eqs.(5\&9)} = (\lvert M_Y^R \rvert \times \varepsilon_Y^{Eq.5}) = (\lvert 1-[{}^WR/{}^SR] \rvert) = 0.0994\%^{*2}$ $(0.3642\%)^{*3}$ | 8.9671 |
| Eq. 6 $^{*4}$ | $M_X^Y = \left(\frac{\partial Y}{\partial X}\right)\frac{X}{Y} = \frac{X\,(C1+1)}{p} = \frac{{}^SR-{}^WR}{{}^SR-{}^DR} = -0.891^{*2}; (-3.266)^{*3}$<br>$M_{Z1}^Y = \left(\frac{\partial Y}{\partial Z1}\right)\frac{Z1}{Y} = -\frac{Z1\,(X+1)\,(C1+1)}{p\,(Z1+1)} = \frac{{}^SR({}^WR-{}^{A1}R)}{{}^{A1}R({}^SR-{}^DR)} = 1.338^{*2};$<br>$(3.582)^{*3}$; $[0.0706^{*2}; (2.687)^{*3}]^{*5}$ | $[UF]_Y^{Eq.6} = (\lvert M_X^Y \rvert + \lvert M_{Z1}^Y \rvert) = 2.23^{*2}$; $(6.85)^{*3}$; $[0.962^{*2}; (5.95)^{*3}]^{*5}$ | $\varepsilon_Y^{Eq.6} = ([UF]_Y^{Eq.6} \times {}^Gu) = 2.23\%^{*2}$ $(6.85\%)^{*3}$; $[0.9617\%^{*2}; (5.953\%^{*3})]^{*5}$ | ${}^S\varepsilon_R^{Eqs.(6\&9)} = (\lvert M_Y^R \rvert \times \varepsilon_Y^{Eq.6}) = 0.2486\%^{*2}$ $(0.7636\%)^{*3}$; $[0.1072\%^{*2}; (0.6638\%)^{*3}]^{*5}$ | 8.9671 |
| Eq. 7 $^{*6}$ | $M_X^Y = \left(\frac{\partial Y}{\partial X}\right)\frac{X}{Y} = \frac{X\,(C1-C2)}{q} = \frac{{}^SR-{}^WR}{{}^SR-{}^DR} = -0.891^{*2}; (-3.266)^{*3}$<br>$M_{Z1}^Y = \left(\frac{\partial Y}{\partial Z1}\right)\frac{Z1}{Y} = \frac{Z1(X+1)\,(C2-C1)}{q\,(Z1-Z2)} = \frac{{}^SR({}^WR-{}^{A1}R)}{({}^{A1}R-{}^{A2}R)({}^SR-{}^DR)} = 9.392^{*2};$<br>$(25.137)^{*3}$<br>$M_{Z2}^Y = \left(\frac{\partial Y}{\partial Z2}\right)\frac{Z2}{Y} = \frac{Z2(X+1)\,(C1-C2)}{q\,(Z1-Z2)} = \frac{{}^SR({}^{A2}R-{}^WR)}{({}^{A1}R-{}^{A2}R)({}^SR-{}^DR)} = -0.425^{*2};$<br>$(16.170)^{*3}$ | $[UF]_Y^{Eq.7} = (\lvert M_X^Y \rvert + \lvert M_{Z1}^Y \rvert + \lvert M_{Z2}^Y \rvert) = 10.7^{*2}$; $(44.6)^{*3}$ | $\varepsilon_Y^{Eq.7} = ([UF]_Y^{Eq.7} \times {}^Gu) = 10.71\%^{*2}$ $(44.57\%)^{*3}$ | ${}^S\varepsilon_R^{Eqs.(7\&9)} = (\lvert M_Y^R \rvert \times \varepsilon_Y^{Eq.7}) = 1.194\%^{*2}$ $(4.97\%)^{*3}$ | 8.9671 |

***1:** ${}^S\varepsilon_R = (\lvert M_Y^R \rvert \times \varepsilon_Y) = ([UF]_R \times {}^Gu)$, with: $[UF]_R = (\lvert M_Y^R \rvert \times ([UF]_Y) = (\lvert ({}^SR - {}^DR)/{}^SR \rvert \times ([UF]_Y) = (0.11152 \times [UF]_Y)$

***2:** This data refers to the measurement-reference "***W***" as GISP (cf. the text and Table 1).

***3:** This value corresponds to SLAP as "***W***" (cf. the text and Table 2).

***4:** $p = (X + 1) \times (C1 + 1) - (Z1 + 1)$; cf. [corresponding to Eq. 6] the expression of "$M_i^Y$".

***5:** These values refer to NBS-1A (i.e. instead of NBS-1) as "***A1***" (cf. the text and the BLOCK nos. **2a** in Tables 1 & 2).

***6:** $q = (X + 1) \times (C1 - C2) - (Z1 - Z2)$; cf. [corresponding to Eq. 7] the expression of "$M_i^Y$".

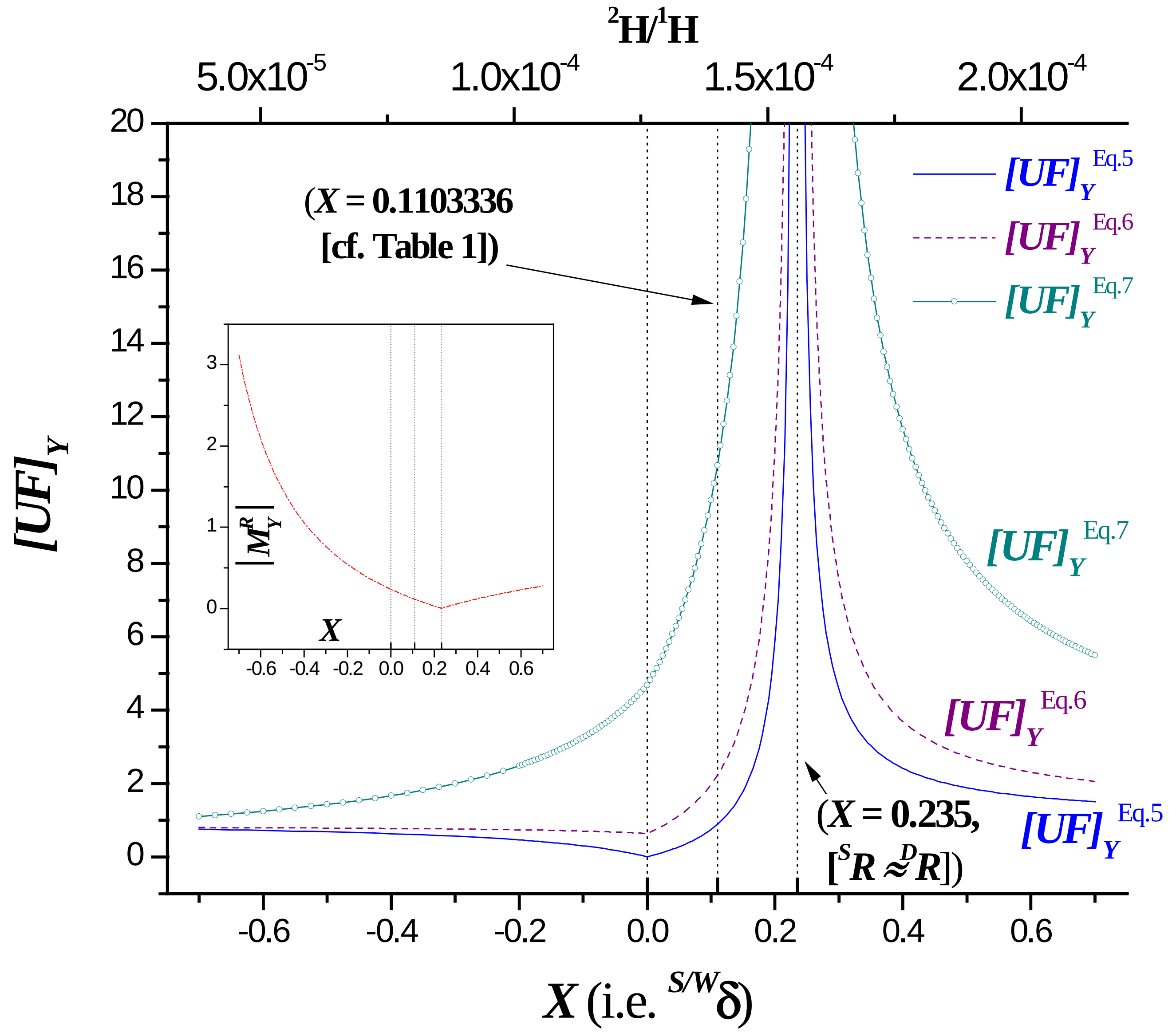

**Figure 1.** Variation of method-specific-$[UF]_Y$ as a function of sample $S$ ($^2$H/$^1$H abundance ratio)

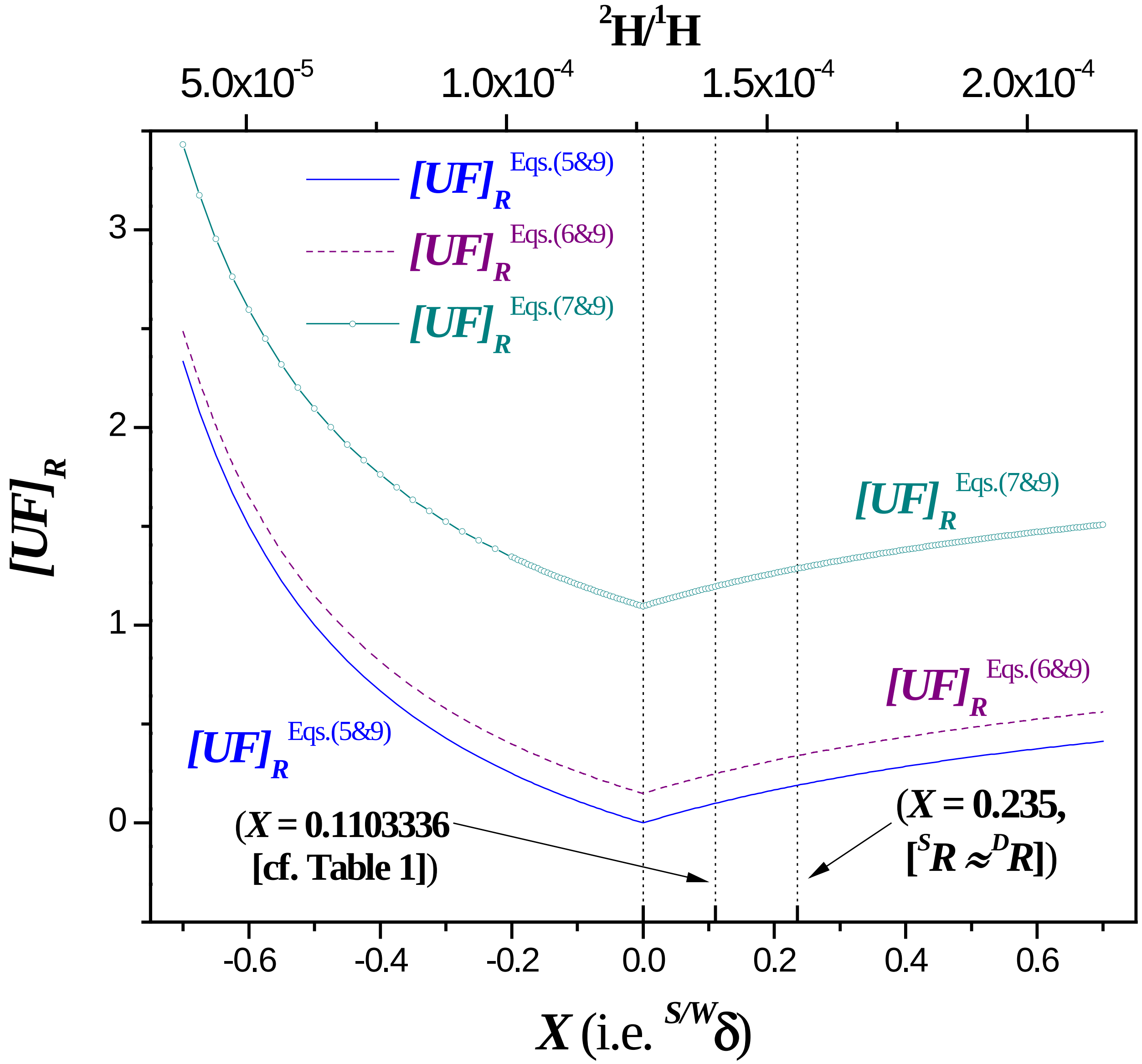

**Figure 2.** Plot of $[UF]_R$ against sample *S* ($^2$H/$^1$H abundance ratio)